\documentclass[%
 reprint,
 amsmath,amssymb,
 aps,
]{revtex4-2}

\usepackage{graphicx}
\usepackage{xcolor}
\usepackage{epigraph}
\usepackage{dcolumn}
\usepackage{bm}

\begin{document}
\title{Maximum Expected Reward Lines If Non Specialist (for darts)}
\author{Merlin Füllgraf}
\email{mfuellgraf@uos.de}
\author{Jochen Gemmer}%
\affiliation{
 Institute for Physics, University Osnabrück, 49076 Osnabrück
}%
\date{April 01, 2024}

\begin{abstract}
Owning up to the authors' occasional mixed performances when playing darts we follow up on the ingenious work to determine the optimal aim to score high by Tibshirani et.\ al \cite{statistician} and expand on maximal expected reward lines assuming spread profiles beyond the standard case of an isotropic distribution by investigating situations in which a players' aiming quality differs in horizontal and vertical direction, arguably representing more realistic profiles. To render the theoretical findings applicable, we determine regions $\Gamma$ on the dart board that may serve as an aid to players in order to maximise their score depending on whether said areas are hit sufficiently often. Lastly we compare all three strategies with the simple approaches to always aim for triple 20 or bulls eye respectively, probing their usefulness.

\end{abstract}
\epigraph{\hfill \textsc{April First Letters 2024}}{}
\maketitle

\section{\label{sec:level1}Introduction}
\begin{figure}[b]
	 \includegraphics[width=0.3\textwidth]{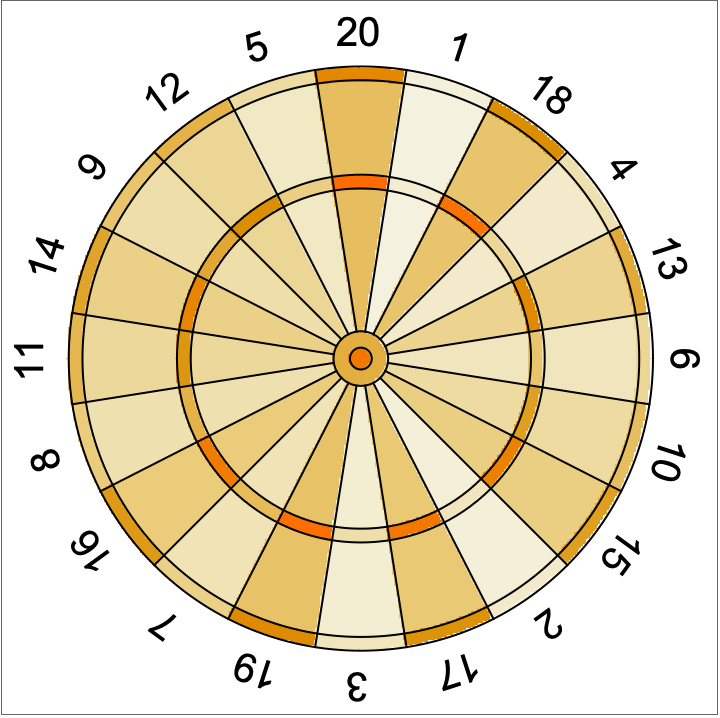}
  \caption{\label{fig:board}Representation of the board $\mathcal{B}$.
  }
\end{figure}

Everyone who has tried to throw darts has found out that hitting the intended field on the board turns out to be substantially more difficult than it seems on TV. Moreover due to the design of the dart board, see Fig. \ref{fig:board}, it is not evident whether there are regions where, given a certain imprecision of the throw, you can expect higher scores than in others.

Over the years the sport also caught the attention of the scientific community. Work on this topic ranges from optimal strategies to finish the game as quickly as possible \cite{optimal,Haugh2020PlayLT,choking,wall} over possible rearrangements of the board itself \cite{rearranging1,rearranging2,rearranging3,rearranging4,rearranging5} to the influence the players' accuracy on the score \cite{statistician,statistician2,score1,score2,score2b,score3,score4,score5} to name a few. 

We model the imprecisions in the throw of the dartist by different kinds of normal distributions
and determine the position for which to aim in order to maximise the expected reward as well as tracking this position along the increasing width of the spread underlying the throw, as employed in \cite{statistician}. 

Moreover we formulate an illustrative criterion serving as an aid to dartists on whether an adjustment of the aim is favourable or not, given that the quality of the throw in terms of the spread does not change across the boards by introducing critical regions $\Gamma$.

Finally we conclude by investigating whether the strategy put forward by the maximal expected reward lines (MERL) actually yields a significant improvement as opposed to always aiming at the highest field on the board, the triple-20 (T20), or simply targeting the middle of the board, the bulls eye (BE). 
\section{The model}
In order to investigate how different spreads influence the expected reward of the throw we first need to concretise how we model the throw. Assuming a normal distribution in the plane of the board which shall be parametrised by $(x,y)$, we compute the reward function as the convolution 
of the kernel given by
\begin{align}
    g_{\left(\sigma_x,\sigma_y\right)}\left(x,y\right)=\frac{1}{2\pi \sigma_x \sigma_y}\exp\left[-\frac{1}{2}\left(\frac{(x-x_0)^2}{\sigma_x^2}+\frac{(y-y_0)^2}{\sigma_y^2}\right)\right]
\end{align}
with the board $\mathcal{B}(x,y)$. The board $\mathcal{B}$ itself is a scalar function whose values are specified in the darts rulebook \cite{pdc,dimension}. In Fig.\ \ref{fig:board} we depict the board $\mathcal{B}$ which for our numerical purposes was set up as a $401\times401$ matrix. 
Modelling the deviation within a throw allows for various free parameters. For example one might argue that some players are better at gauging their throw in horizontal or vertical direction respectively which would result in fixed ratios for $\sigma_x/\sigma_y$ or that the distinction between left- and right-handed dartists causes tilted spread patterns \cite{haugh2024empirical,score4}. In this work we study the three simplest cases by revisiting the \textit{isotropic} case \cite{statistician,wall}, where the spread in the horizontal and vertical direction are assumed to be equal, i.e.\ $\sigma_x=\sigma_y$, and investigating the case where a player is better at adjusting in one of these directions.

As argued in \cite{statistician} a smaller spread in horizontal direction is commonly expected due to the throwing motion. Hence we expect this scenario to possibly serve as a toy model for hobby dartists like us.

\begin{figure}[t]
	 \includegraphics[width=0.3\textwidth]{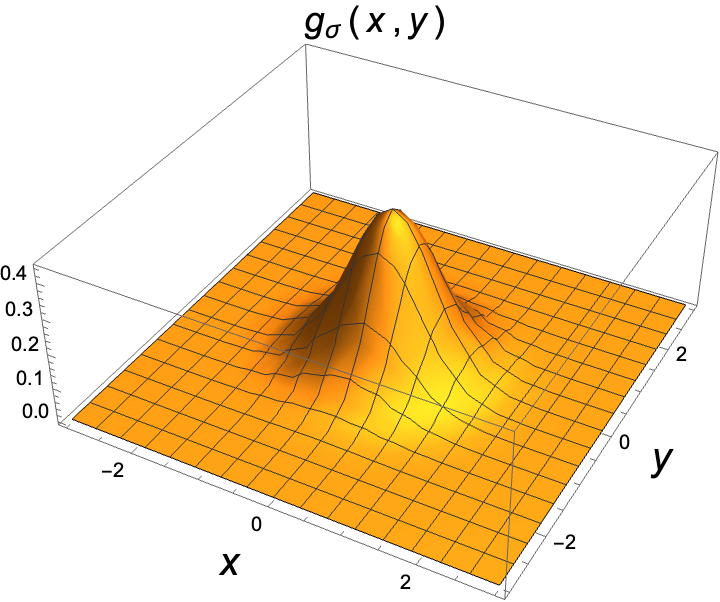}
  	 \includegraphics[width=0.3\textwidth]{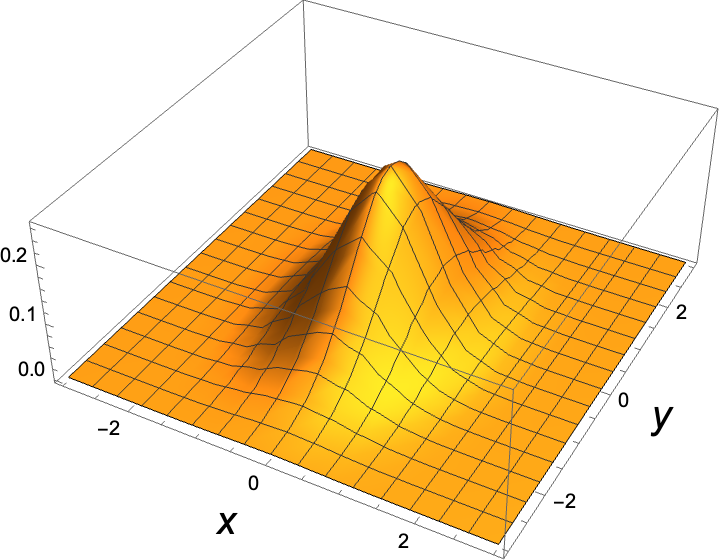}
  \caption{\label{fig:gaussian}\textit{Top:} Gaussian spread profile for equal standard deviations $\sigma_x=\sigma_y$. \textit{Bottom:} Spread where $\sigma_y=2\sigma_x$, i.e.\ modeling the vertical case.
  }
\end{figure}

To this end we study normal distributions where one standard deviation is twice as large as the other one. We dub these cases \textit{horizontal} and \textit{vertical}, indicating in which direction the spread is larger, i.e.\ $\sigma_x=2\sigma_y$ or $\sigma_y=2\sigma_x$ for \textit{horizontal} and \textit{vertical} respectively. 
Sketches for these kernels are depicted in Fig.\ \ref{fig:gaussian}.

Having set up the machinery the expectation value landscape (EVL) is computed $\mathcal{J}_\sigma$ via
\begin{align}\begin{split}
        \mathcal{J}_\sigma(x,y)&:=\left(\mathcal{B}\ast g_{\sigma}\right)(x,y)\\&=\int_{\mathbb{R}^2}\text{d}x^\prime\text{d}y^\prime \mathcal{B}(x^\prime,y^\prime)g_{\sigma}(x-x^\prime,y-y^\prime)
\end{split}
\end{align}
 for various spread widths $\sigma$.
 \section{Results}
\begin{figure}
	 \includegraphics[width=0.2\textwidth]{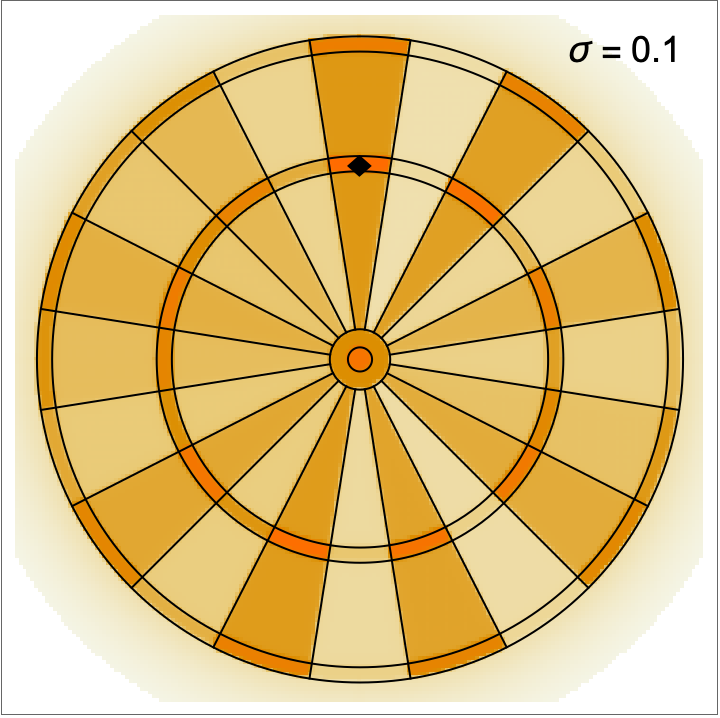}
  	 \includegraphics[width=0.2\textwidth]{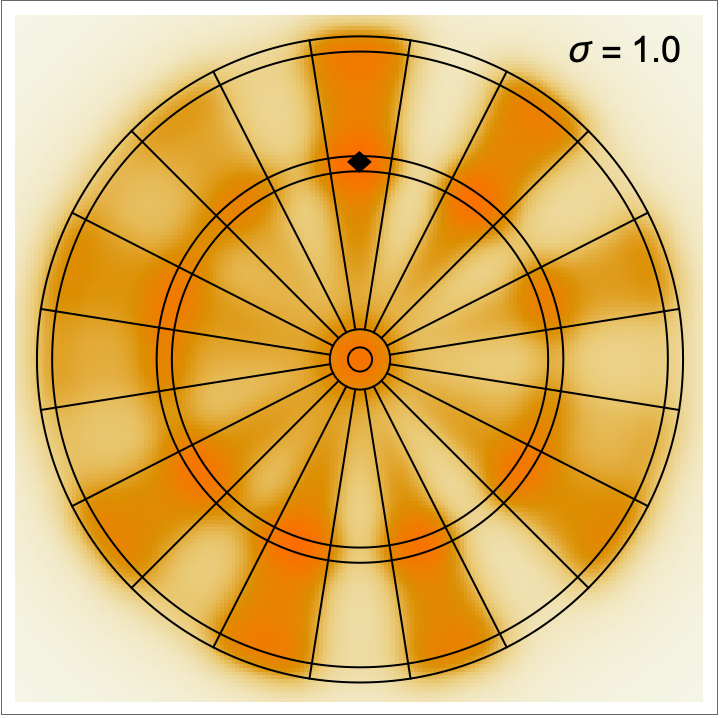}\\
     \includegraphics[width=0.2\textwidth]{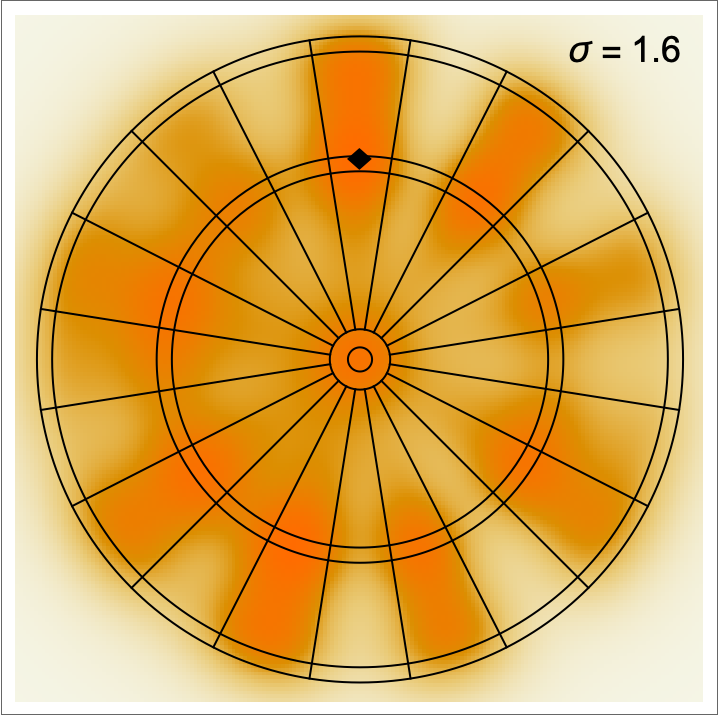}
  	 \includegraphics[width=0.2\textwidth]{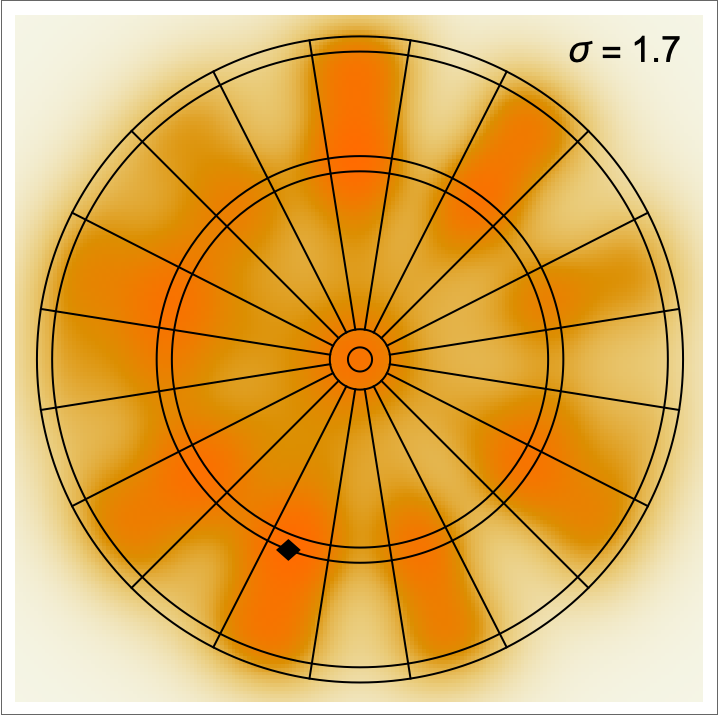} \\
    \includegraphics[width=0.2\textwidth]{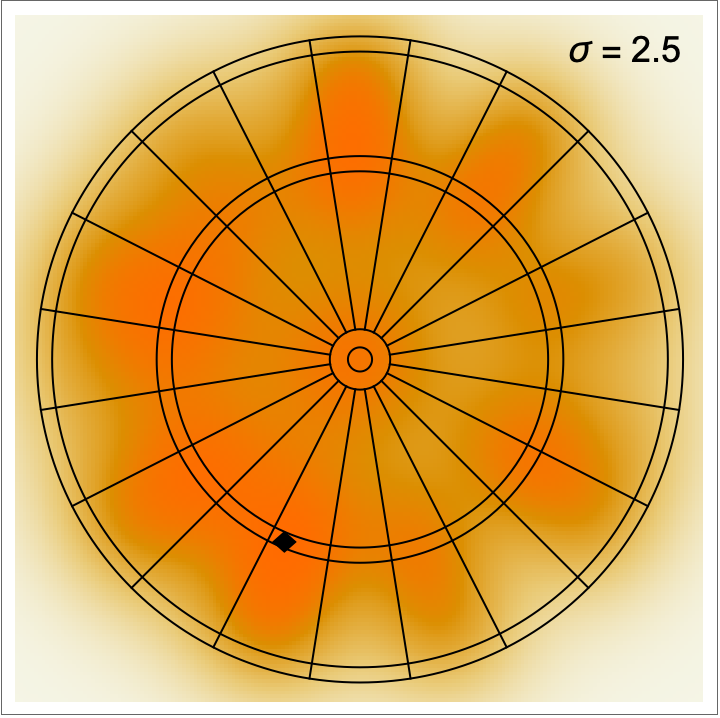}
  	 \includegraphics[width=0.2\textwidth]{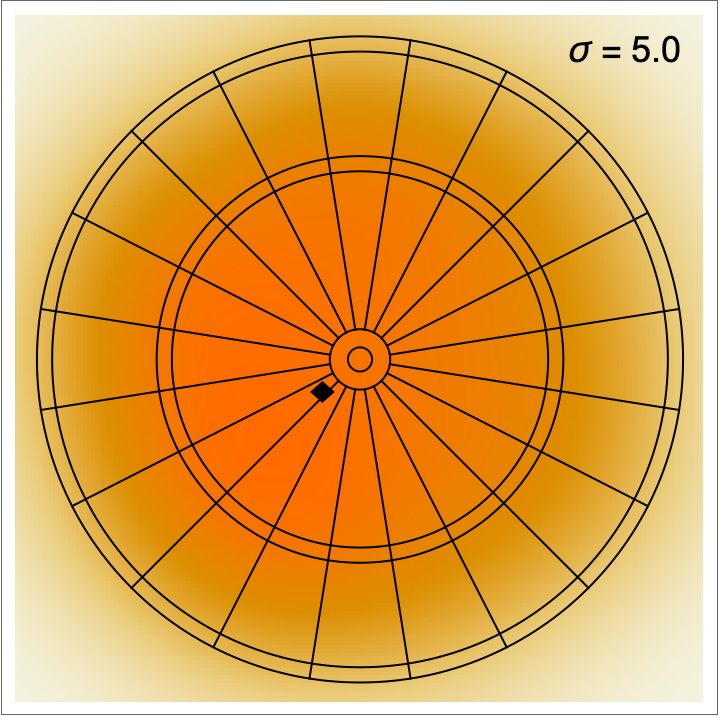}
  \caption{\label{fig:process}Expectation value landscapes (EVL) $\mathcal{J}_\sigma$ with the position of its respective maximum for different spreads $\sigma$ in the isotropic case. Upon exceeding $\sigma_{\text{crit}}=1.6$cm a jump of the CHEN is observed.}
\end{figure}

\begin{figure}
	 \includegraphics[width=0.35\textwidth]{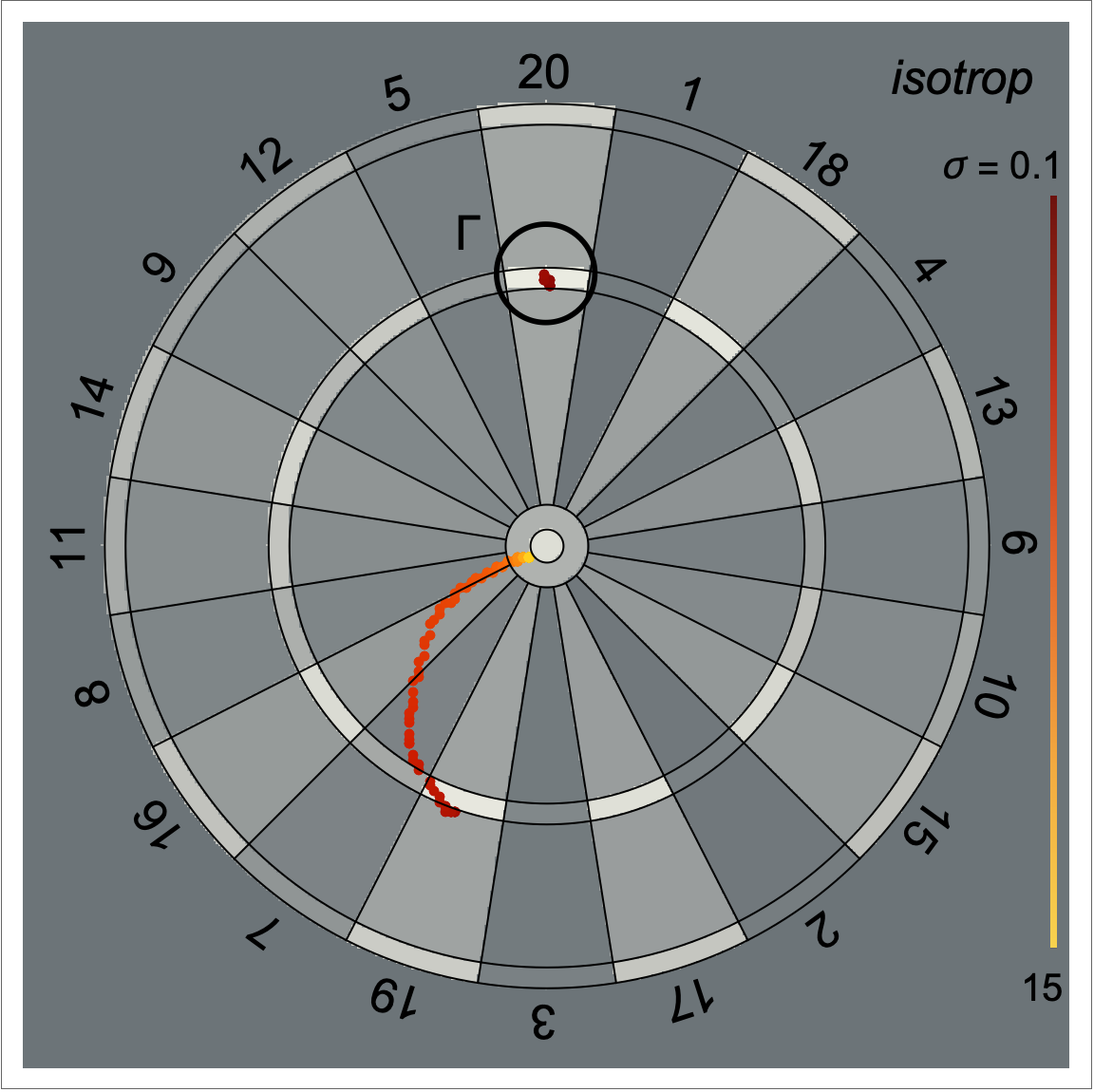}
  \caption{\label{fig:routeiso}Maximum Expected Reward Line for the isotropic case with $\sigma_x=\sigma_y$.
  }
\end{figure}

\begin{figure}
	 \includegraphics[width=0.225\textwidth]{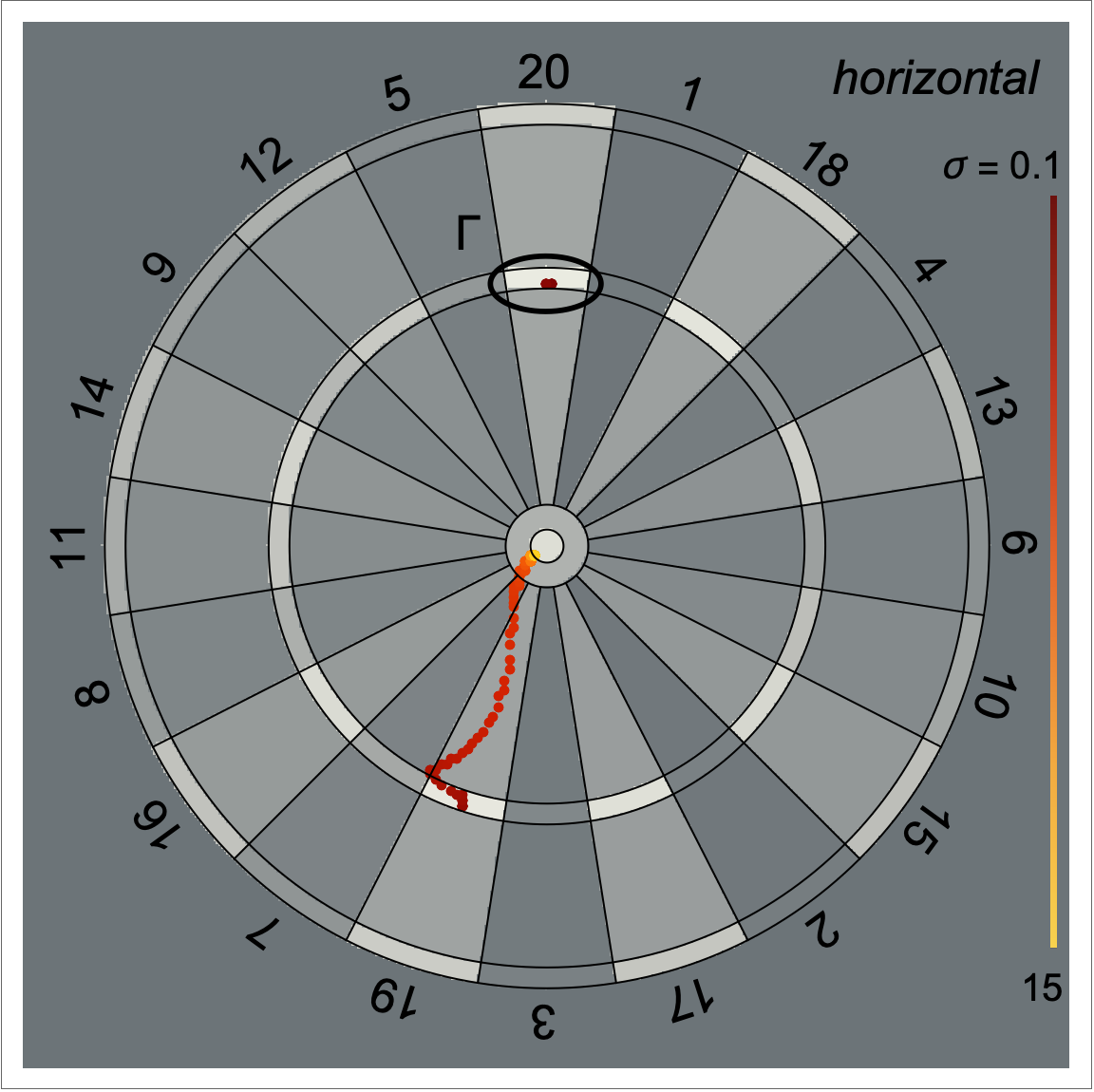}
	 \includegraphics[width=0.225\textwidth]{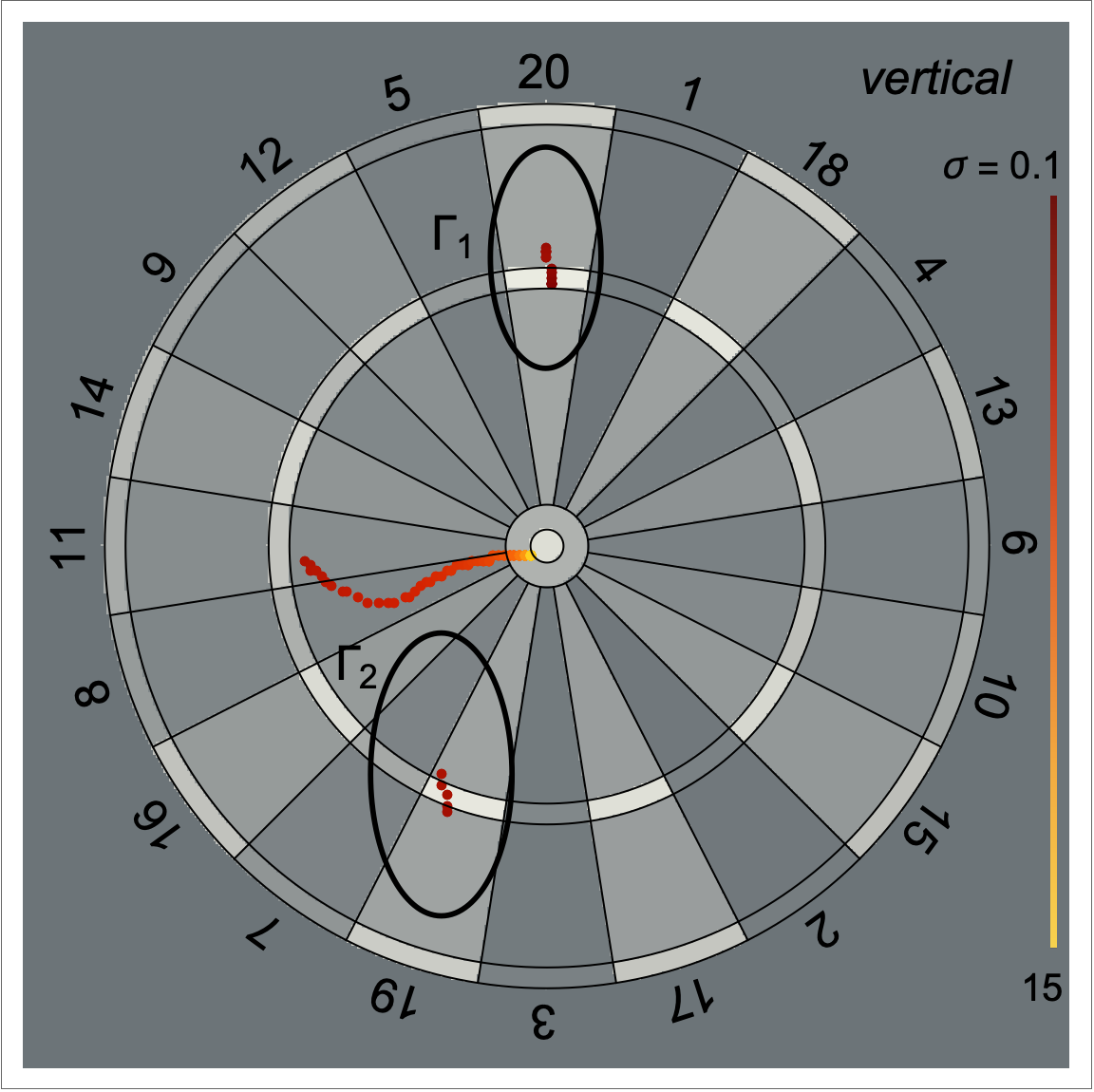}
  \caption{\label{fig:route}Maximum Expected Reward Lines for the horizontal (left, $\sigma_x/\sigma_y=2$) and vertical (right, $\sigma_x/\sigma_y=1/2$) respectively.
  }
\end{figure}

 We compute the expectation value landscape $\mathcal{J}_\sigma$ for spreads ranging from $\sigma=0.1$cm to $15$cm. For the anisotropic cases we write $\sigma=\min\{\sigma_x,\sigma_y\}$. For small $\sigma$ the EVL essentially coincides with the mere board $\mathcal{B}$ whereas it smears out as the spread in the throw grows, i.e.\ the aim of the player gets worse. This process is shown in Fig.\ \ref{fig:process} at the example of the isotropic case. 
For professional players who throw very precisely, i.e.\ have a very small spread, it is reasonable to expect that the highest field of the board (triple 20) will also result in the highest outcome. For newcomers who have a hard time hitting the board, it is presumably wisest to aim for the middle of the board, i.e.\ the bulls eye, in an effort to make sure to score at all. 
Tracking the position of the maxima of each EVL on the board along $\sigma$ we determine the maximum expected reward line (MERL) for each scenario, retrieving in Fig.\ \ref{fig:routeiso} the findings first made in \cite{statistician} for the isotropic case. In Fig.\ \ref{fig:route} we find the MERLs for the horizontal and vertical case respectively.

In accordance with the previous reasoning the two extremes for highly accurate players as well as for absolute greenhorns lie on the triple-20 and close to the bulls eye respectively. 

We find that with increasing spread of the profiles every case exhibits discontinuities in the respective MERL, associated to jumps of the critical highest-expectation value neighbourhoods (CHEN). Before delving into the question how the abstract quantity $\sigma$ effectively can be used to advocate for different aims, we discuss the qualitative structure of the MERLs. For the sake of completeness we include the isotropic scenario here as well.

For the isotropic case we see upon exceeding $\sigma_\text{crit}=1.6$cm the maximum of the respective EVL jumping from within the triple-20 segment to triple-19 before following an arc across the 7 and 16 heading towards the bulls eye. In the horizontal case the picture is similar with the MERL starting within the triple-20 before jumping to the triple-19 at $\sigma_\text{crit}=0.9$cm. Upon then the MERL follows on in the direction of the 7 before turning back to the centre of the 19 segment while steadily moving towards the bulls eye. The MERL of the vertical case exhibits the most features. It starts of at the triple-20 from which it jumps to the triple-19 at $\sigma_{\text{crit,1}}=1.8$cm. Moving slightly upwards it exhibits another jump at $\sigma_\text{crit,2}=2.3$cm ending in the 11 segment. Heading towards the bulls eye it follows an arc through the 8. 

Hence, given that the assumption is true that a common distribution is characterised by a larger spread in vertical than horizontal direction, for most dart players the personal MERL differs qualitatively from the simple picture sketched in Fig.\ \ref{fig:routeiso}.

Without any interpretation, $\sigma$ remains abstract and lacks applicability. To address this apparent shortcoming we define regions $\Gamma$ centered around each point of the MERL satisfying the condition that some fixed percentage of all thrown darts fall into this region, allowing a feasible understanding of jumps of the CHEN. For our purposes we focus on the case that 50$\%$ need to hit $\Gamma$. Moreover we are still free to set the geometry of the region. In consideration of the underlying probability distribution we set up $\Gamma$ as ellipses, i.e.\ we determine $a,b$ such that
\begin{align}
    \eta&=\int_\Gamma \text{d}x\ \text{d}y \ g_{\left(\sigma_x,\sigma_y\right)},\quad \eta=\frac{1}{2},
    \\ \text{with}\ \Gamma&=\left\{(x,y)\in\mathbb{R}^2\ \vline\ \frac{x^2}{a^2}+\frac{y^2}{b^2}=1\right\}.
\end{align}
For the isotropic case we require $\Gamma$ to also be radially symmetrical, i.e.\ a circle with radius $a$. For the other spread profiles we set $a=2b$ and $b=2a$ for the horizontal and vertical cases respectively. Professional dart players however can aim so well that for them the shape of $\Gamma$ depends on the field they aim at \cite{haugh2024empirical}. Therefore when targeting fields at the top or bottom of the board, e.g.\ D20 the profile is similar to the horizontal spread, whereas at the very left and right of the board, e.g.\ D11 the spread is more along the vertical direction. Nevertheless, speaking of mere mortals, we for the entirety of this article stick to uniform distributions across the board.
Denoting the small half axes by $r_\text{m}:=\min\{a,b\}$, we find $r_\text{m}\approx1.8$cm for the isotropic case, $r_\text{m}\approx1.06$cm for the horizontal one and for the vertical spread profile $r_\text{m,1}\approx2.12$cm and $r_\text{m,2}\approx2.71$cm. The respective regions $\Gamma$ are depicted in Fig. \ref{fig:routeiso} and \ref{fig:route} and visualize how well the players' aim at least needs to be before switching according to the jump in the MERL. Following the assumption put forward in \cite{statistician} that a vertical profile is common due to the motion of the throw, from $\Gamma_1$ shown in Fig.\ \ref{fig:route} and $\sigma_{\text{crit},1}$ we can infer that with this kind of throw distribution aiming for T20 is advised the longest (in terms of $\sigma$) as the critical value for the jump as well as the region of $\Gamma$ are the largest. 

\begin{figure}[h]
	 \includegraphics[width=0.4\textwidth]{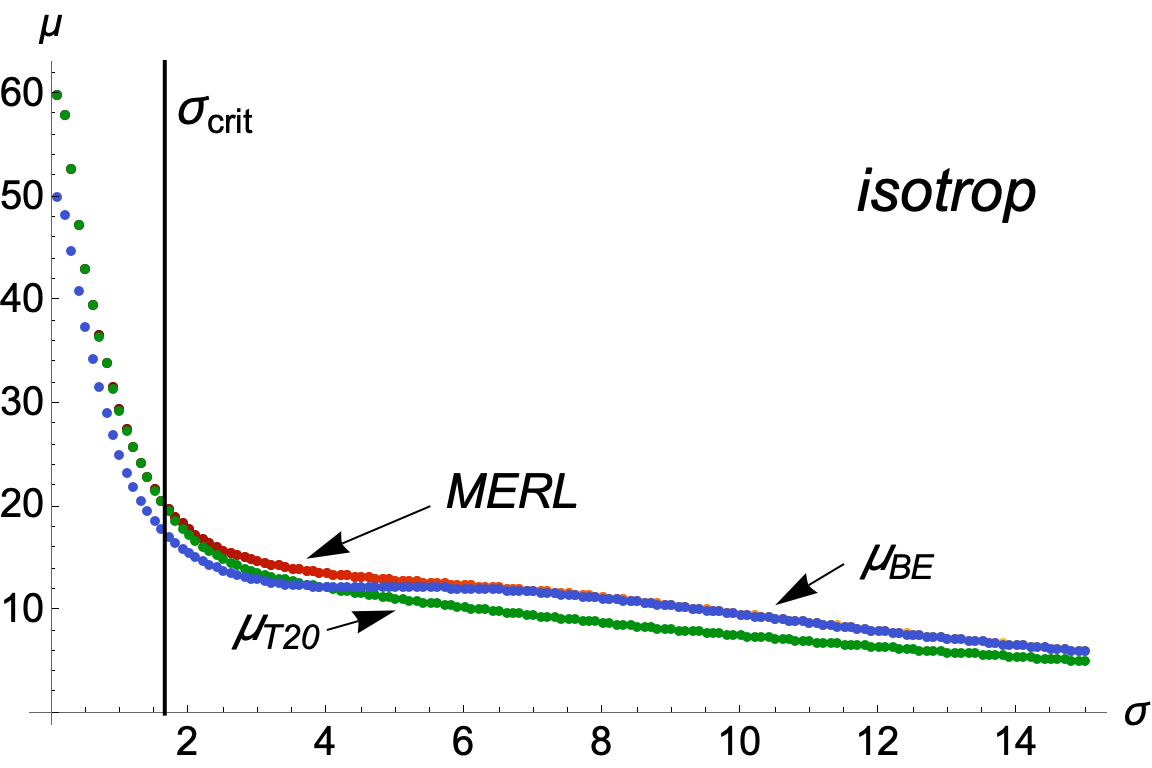}
	 \includegraphics[width=0.235\textwidth]{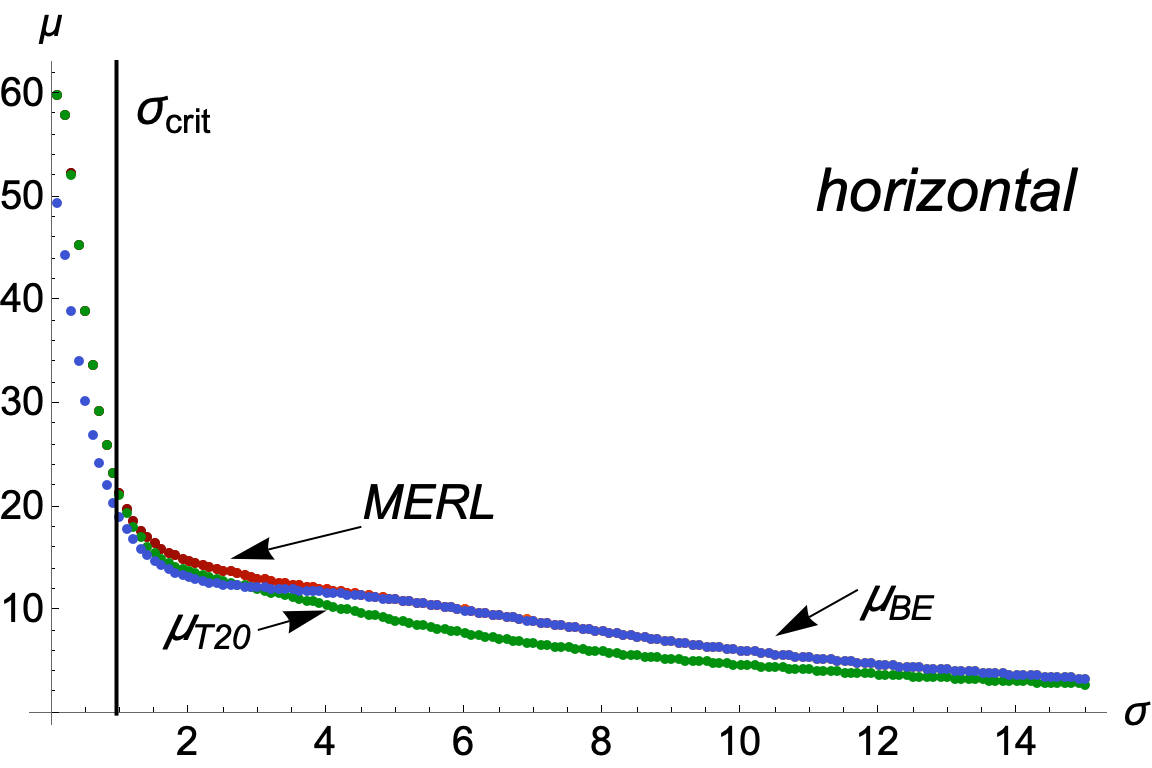}
   \includegraphics[width=0.235\textwidth]{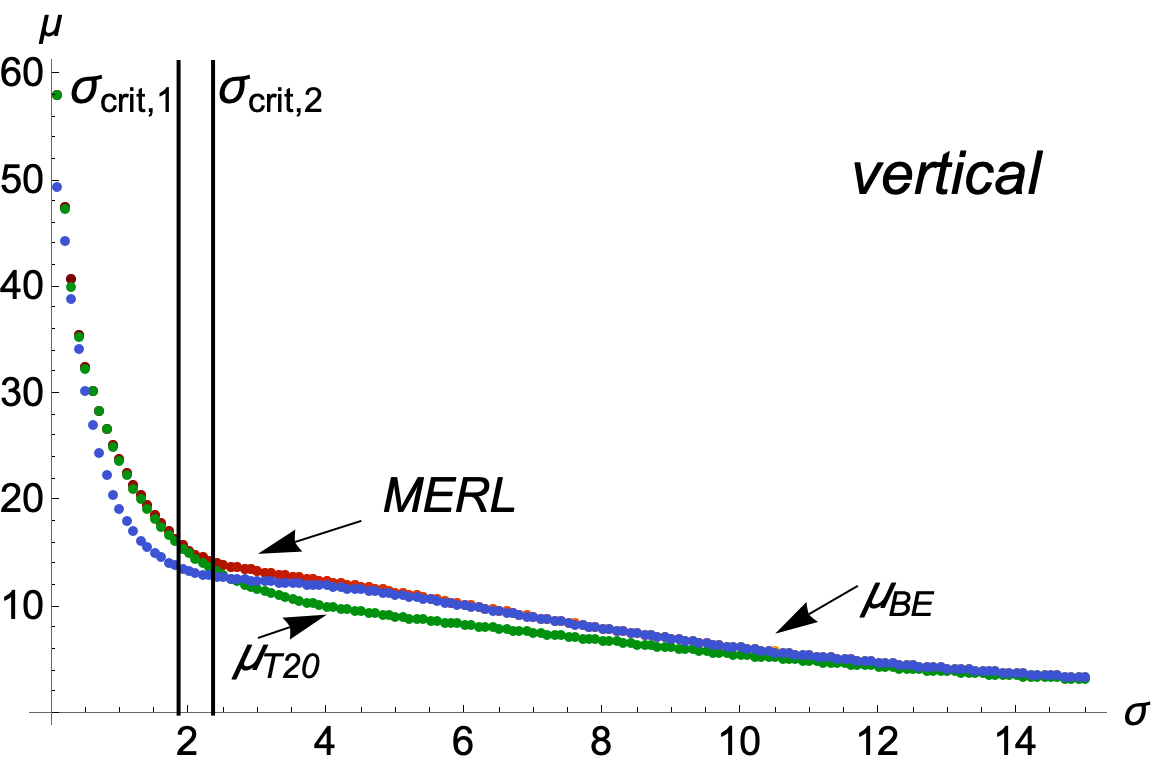}
   \caption{\label{fig:mu_line}Expected reward according to the MERL (red) and upon always aiming for triple-20 ($\mu_{\text{T20}}$, green) and bulls eye ($\mu_{\text{BE}}$, blue). \textit{Top}: Isotropic case. \textit{Bottom left}: Horizontal and \textit{bottom right}: vertical case.}
\end{figure}

Lastly we study whether following the MERL can actually give rise to a significant improvement concerning the expected outcome of the throw opposed to simply always aiming at triple-20 which yields an expectation value we denote $\mu_\text{T20}$. With $\mu_{\text{BE}}$ we denote the expectation value for the most cautious strategy of always aiming the bulls eye. We depict the expected return $\mu$ following the different strategies in Fig. \ref{fig:mu_line}, recovering the behaviour for the isotropic case found in \cite{wall}.
Until the first jump the MERL is in the triple-20 section and both strategies coincide. After that the lines fan out, the MERL turns to higher rewards than simply aiming for triple-20 up to the point where the spread is so broad that hitting the board in the first place becomes a challenge and the precise aim becomes subordinate. In this process the MERL and the line corresponding to bulls eye-strategy converge.

\begin{figure}[]
	 \includegraphics[width=0.235\textwidth]{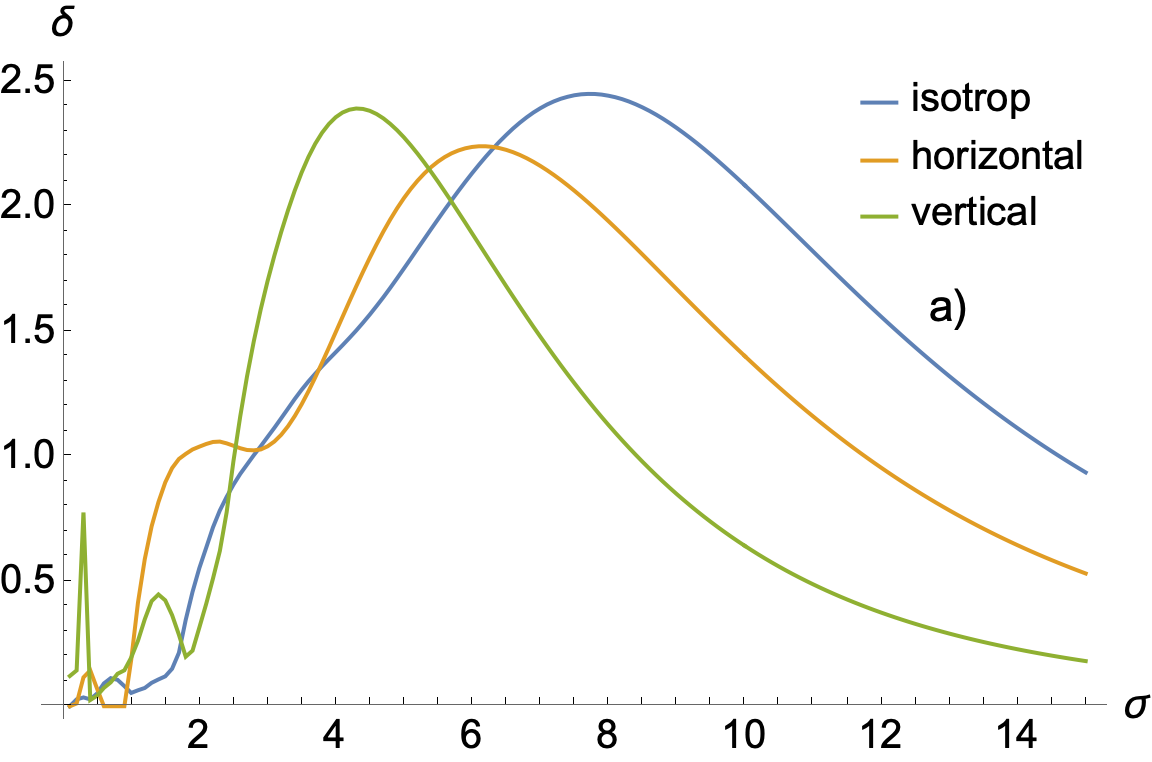}
  \includegraphics[width=0.235\textwidth]{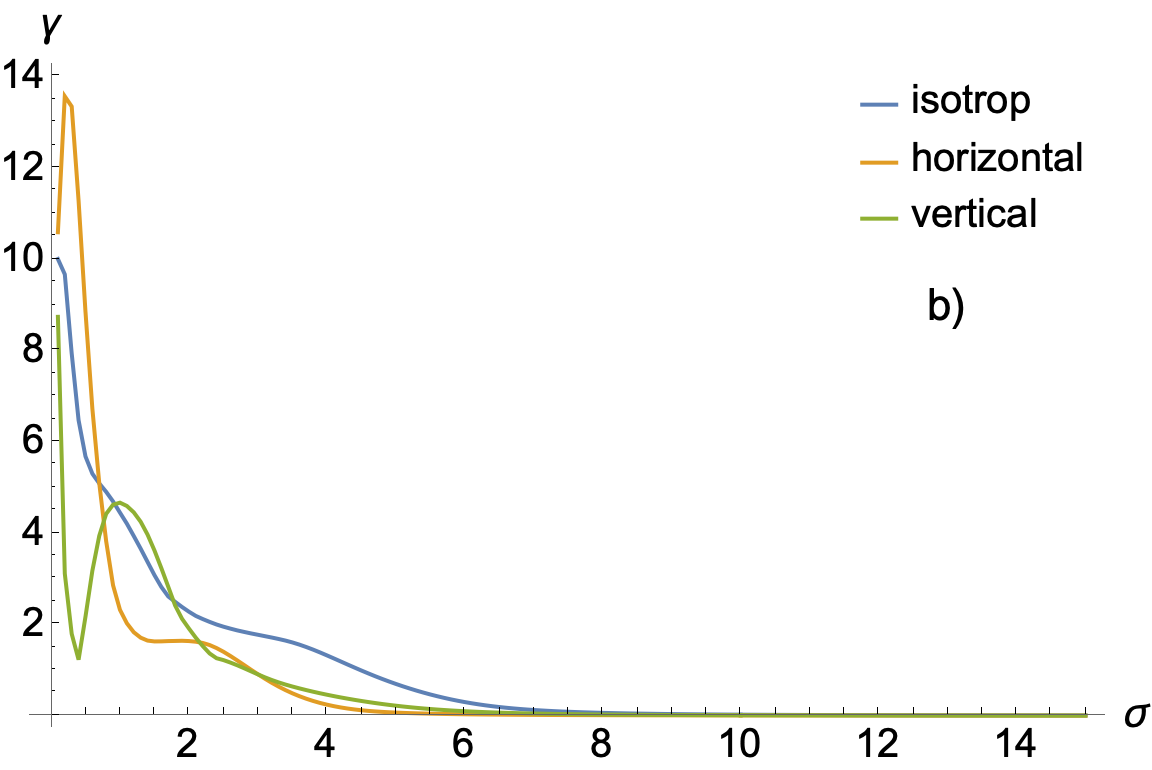}
  \caption{\label{fig:advantage}\textbf{a)} Difference $\delta$ between the expectation values according to the MERL and unwavering aiming for triple-20 along the spread parameter $\sigma$. \textbf{b)} Advantage $\gamma$ of the strategy according to the MERL over always targeting the bulls eye in dependence of $\sigma$.}
\end{figure}

Measuring the difference $\delta$ between following the MERL and consistently throwing at triple-20 we find in Fig. \ref{fig:advantage}\ a) that adjusting the strategy accordingly can give rise to up to $7.5$ points per three darts. 

Studying the advantage $\gamma$ of the MERL over $\mu_{\text{BE}}$ (see Fig.\ \ref{fig:advantage}\ b)) we find that especially at low $\sigma$ the strategy to give rise to higher rewards in the order of 10 per throw across all three considered spread models. Also at intermediate width $\sigma$ from 2cm to 4cm, following the MERL promises an increased reward of up to 6 points per three darts. However for a spread width $\sigma\approx6$cm the inaccuracies are so large that benefits from following a strategy such as the one put forward here become negligible as the MERL and $\mu_{\text{BE}}(\sigma)$ converge.

\section{Conclusion and outlook}
In this paper we studied how the ideal aim on a dart board changes when the spread of the underlying throw grows for anisotropic throw distributions in horizontal and vertical direction. We found that the resulting ideal aims gives rise to a discontinuous curve on the dart board. Assuming that for mediocre players due to the motion of the throw, the spread profile is similar to the vertical case, we demonstrate that the optimal aiming strategy differs substantially from the isotropic scenario (see Figs.\ \ref{fig:routeiso} \&\ \ref{fig:route}), possibly giving rise to a rethink of the personal aiming strategy.

To capture whether the strategy proposed by these lines is favourable and to make the numerics accessible we introduced areas $\Gamma$ that can serve as a useful guide to dart players to determine whether changes according to these strategies can be wise. Comparing with the approach of always aiming at T20 we find an increased score of up to 7.5 points per turn. Nevertheless we see that starting at intermediate spreads $\sigma$ the advantage of a specific strategy starts to become negligible. 

However, we wish to stress that all findings are based on the underlying assumptions that the spread can be described by a normal distribution as well as that the spread is identical for all aims, i.e.\ there are no regions in with the throw is more accurate than in others.
Hence we encourage every interested reader to go ahead, grab some friends, hit a nearby pub and probe the model themselves. In this sense: Game on!

\section{Acknowledgements}
We would like to thank Robin Steinigeweg for fruitful discussions on the numerical aspects of this project. Special thanks also go to Mariel Kempa, Markus Kraft and Jiaozi Wang for helping us to assess the content of this paper in an empirical manner in the margin of the DPG spring meeting 2024 in Berlin.
MF would also like to thank Daniel Westerfeld for interesting discussions on darts in general as well as Simon Sperlich and Felix Schäfer for insightful comments on an earlier version of this project.\\
\textit{Note added:} After the first version was published on the arXiv we were made aware of \cite{statistician}.
\bibliography{darts_v2.bib}
\end{document}